\documentclass[conference]{IEEEtran}
\ifCLASSINFOpdf
\else
\fi

\usepackage{subcaption}
\usepackage{graphicx}
\usepackage{xcolor}

\newif\ifcolormarker
\colormarkerfalse

\newcommand{\bbm}[1]{\ifcolormarker
\textcolor{blue}{#1}%
\else
#1%
\fi } 

\begin{document}
%
\title{Towards Real-Time Routing Optimization with Deep Reinforcement Learning: Open Challenges}


\author{\IEEEauthorblockN{Paul~Almasan\IEEEauthorrefmark{1},
José~Suárez-Varela\IEEEauthorrefmark{1},
Bo~Wu\IEEEauthorrefmark{2},
Shihan~Xiao\IEEEauthorrefmark{2},
Pere~Barlet-Ros\IEEEauthorrefmark{1}, 
Albert~Cabellos-Aparicio\IEEEauthorrefmark{1}}
\IEEEauthorblockA{\IEEEauthorrefmark{1}Barcelona Neural Networking Center, 
Universitat Politècnica de Catalunya, Spain\\
}
\IEEEauthorblockA{\IEEEauthorrefmark{2}Network Technology Lab., Huawei Technologies Co.,Ltd.}
\textbf{NOTE}: This is a version of a paper accepted in IEEE HPSR SARNET 2021. Please use the following reference to cite the\\corresponding work: P. Almasan, J. Suárez-Varela, B. Wu, S. Xiao, P. Barlet-Ros and A. Cabellos-Aparicio, "Towards\\Real-Time Routing Optimization with Deep Reinforcement Learning: Open Challenges," 2021 IEEE 22nd International\\Conference on High Performance Switching and Routing (HPSR), 2021, pp. 1-6, doi: 10.1109/HPSR52026.2021.9481864.}


%


\IEEEoverridecommandlockouts
\IEEEpubid{\makebox[\columnwidth]{10.1109/HPSR52026.2021.9481864~\copyright{}2021 IEEE
\hfill} \hspace{\columnsep}\makebox[\columnwidth]{ }}

\maketitle

\begin{abstract}

The digital transformation is pushing the existing network technologies towards new horizons, enabling new applications (e.g., vehicular networks). As a result, the networking community has seen a noticeable increase in the requirements of emerging network applications. One main open challenge is the need to accommodate control systems to highly dynamic network scenarios. Nowadays, existing network optimization technologies do not meet the needed requirements to effectively operate in real time. Some of them are based on hand-crafted heuristics with limited performance and adaptability, while some technologies use optimizers which are often too time-consuming. Recent advances in Deep Reinforcement Learning (DRL) have shown a dramatic improvement in decision-making and automated control problems. Consequently, DRL represents a promising technique to efficiently solve a variety of relevant network optimization problems, such as online routing. In this paper, we explore the use of state-of-the-art DRL technologies for real-time routing optimization and outline some relevant open challenges to achieve production-ready DRL-based solutions.

\end{abstract}


%
\IEEEpeerreviewmaketitle

\section{Introduction}

The digital transformation is driving existing network technologies towards supporting emergent networking paradigms (e.g., vehicular networks~\cite{zeadally2012vehicular}, smart cities~\cite{yaqoob2017enabling}, satellite networks~\cite{alagoz2007exploring}). These modern applications impose new requirements for existing network optimization technologies (e.g.,  adaptability to dynamic scenarios, low latency). In this context, Machine Learning (ML) seems a good candidate to enable real-time operation and allow efficient optimization across dynamic systems \cite{mnih2013playing}. In the present paper, we explore the application of modern ML-based solutions for routing optimization in dynamic network scenarios.  

Traditional solutions for routing optimization can be mainly differentiated in two different categories: $(i)$ Specific-purpose heuristics and \mbox{$(ii)$ Computationally-intensive} mathematical solvers, which intrinsically are not adequate for efficient real-time optimization. Heuristic-based solutions (e.g., shortest path, load balancing) are often sufficiently fast, but they require expert knowledge to be adapted to specific optimization problems, and often do not adapt well to dynamic network scenarios. Solutions based on general-purpose mathematical optimizers (e.g., CPLEX~\cite{cplex})
are often too time-consuming to achieve real-time operation. This pushes the networking community to explore new methods and technologies that can adapt well to the requirements of emerging network applications.

Recent advances in Deep Reinforcement Learning (DRL) showcased that this technology is capable to operate efficiently in complex optimization problems~\cite{mnih2013playing,schrittwieser2020mastering}. However, state-of-the-art DRL-based solutions used in other fields are not directly applicable to modern networks, where it is needed to generalize across scenarios with dynamic changes (e.g., topology, traffic, link failures)~\cite{almasan2019deep}. One key technology that enables DRL to handle network changes is Graph Neural Network (GNN)~\cite{scarselli2008graph}. GNN is a novel family of neural networks specifically intended to operate on graph-structured data. Particularly, these recent neural networks have a strong relational bias over graphs, which results in an unprecedented level of combinatorial generalization over graph-structured data~\cite{battaglia2018relational}. As a result, GNNs achieve outstanding generalization capabilities over networks \cite{almasan2019deep, suarez2019challenging} and, similarly to other deep learning methods, they achieve low execution times -- often sub-second operation in commodity hardware. 

In this paper, we explore the applicability of state-of-the-art solutions based on DRL and GNN to modern networking scenarios with highly dynamic topology changes. Specifically, we design a DRL+GNN architecture for online routing optimization, and discuss the open challenges of this technology. First, we train a GNN-based DRL agent in a single topology, and then we test if the agent is able to operate successfully on different topologies never seen before. This allows us to assess the capabilities of this solution to adapt to fundamental changes in the network topology. Lastly, we discuss the implications and challenges that represents the new paradigm of DRL+GNN.

\section{Related work}
\label{sec:related-work}

Finding the optimal routing configuration given an estimated traffic matrix is a fundamental networking problem, which is known to be NP-hard \cite{xu2011link, hartert2015declarative}. This problem has been largely studied in the past and we outline some of the most relevant works. In DEFO~\cite{hartert2015declarative}, the authors propose a solution that converts high-level optimization goals, indicated by the network operator, into specific routing configurations using constraint programming.  
In \cite{gay2017expect}, the authors propose to use local search, where they sacrifice space exploration to achieve lower execution times. A more recent work \cite{jadin2019cg4sr} leverages the column generation algorithm and dynamic programming to solve Traffic Engineering problems. 

Recently, numerous DRL-based solutions have been proposed to solve network optimization problems. In \cite{xu2018experience}, they propose a generic DRL framework for Traffic Engineering. Their solution is based on a DRL agent that defines the split ratios of a flow over a set of paths. In the field of optical networks, the work \cite{suarez2019feature} proposes an elaborated representation of the network state to help a DRL agent learn to route traffic demands. A more recent work \cite{almasan2019deep} proposes a different approach where the authors combine DRL and GNN \bbm{to optimize the resource allocation in optical circuit-switched networks}. 
The recent work from \cite{sun2020scalable} proposes a scalable DRL-based solution where they use pinning control theory to select critical links in the network and optimize their weights assuming a weighted shortest path routing policy (e.g., OSPF and ECMP).

\section{Routing Optimization scenario}
\label{sec:rout-optim-scenario}

In this paper we explore the challenges when using DRL-based solutions for routing optimization in dynamic network scenarios. Particularly, we consider a network optimization scenario based on Software-Defined Networking (SDN), where the  DRL  agent  (located  in  the  control  plane)  has  a  global view of the current network state, and has to optimize the routing configuration considering the current traffic matrix. This is a complex optimization problem that has been widely studied in the past \cite{hartert2015declarative, jadin2019cg4sr}.

\begin{figure}[!t]
  \centering
  \includegraphics[width=0.98\linewidth]{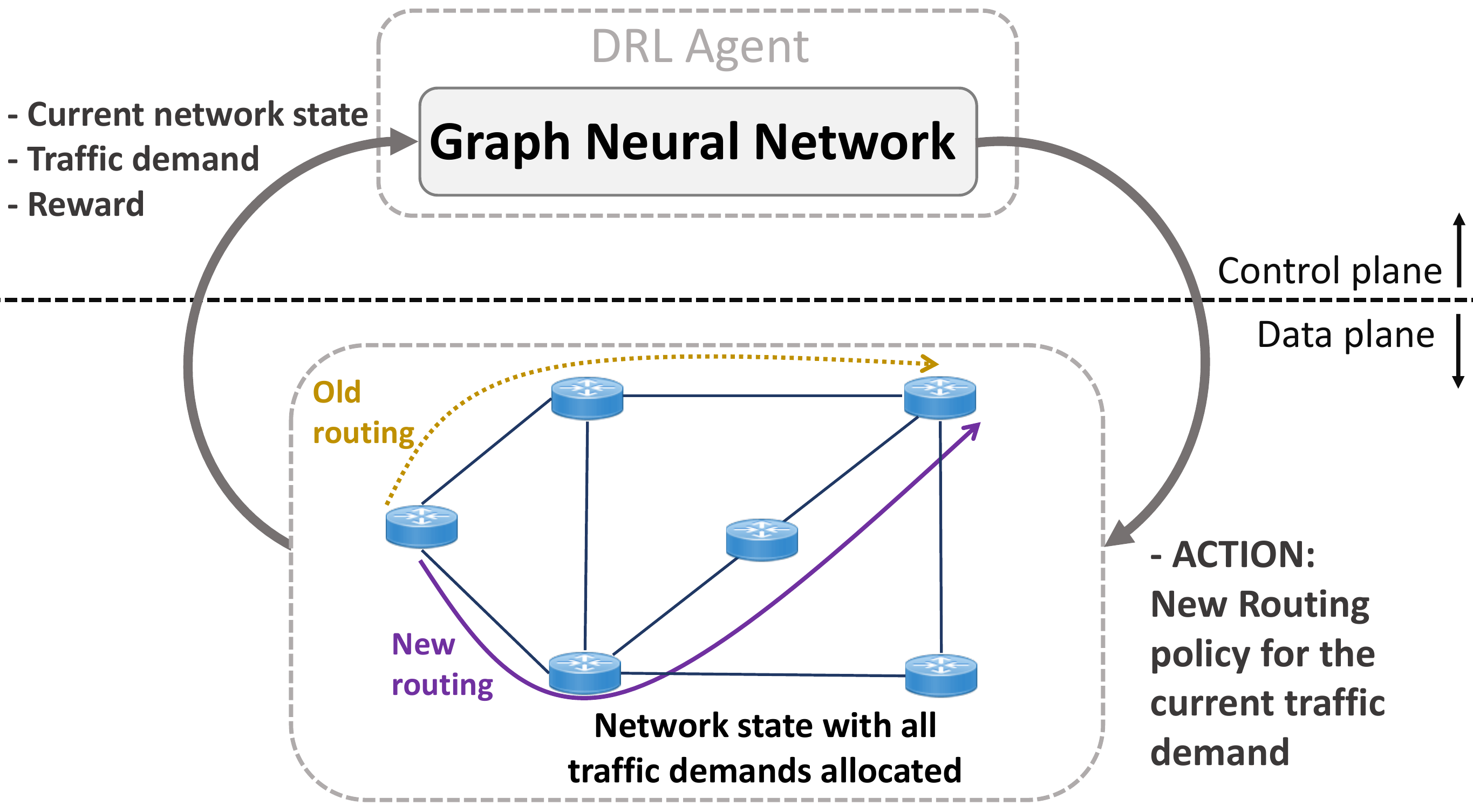}
  \caption{Schematic representation of the DRL agent in the routing optimization scenario.}
  \label{fig:drlEnvMidd}
   \vspace{-0.2cm}
\end{figure}

All the source-destination traffic demands are initially allocated using OSPF routing \cite{moy1998rfc2328} and the DRL agent must find a new routing policy that minimizes the utilization of the most loaded link. This is a common scenario where overlay technologies (e.g., Segment Routing \cite{filsfils2015segment}, LISP \cite{rodriguez2015lisp}) are used in the optimization layer \cite{hartert2015declarative}. Traffic demands are  defined by the tuple $\{src, dst, bw\}$ where \textit{src} and \textit{dst} are the source and destination nodes respectively and \textit{bw} represents the average traffic volume estimated for this demand. 
The DRL agent iterates over all demands and it has to decide the new routing path for each of them. Considering all the possible \textit{src-dst} combinations in the network, there is a total of \textit{N*(N-1)} traffic demands, where \textit{N} is the number of nodes (i.e., one demand per \textit{src-dst} pair). A DRL episode ends when the agent has iterated over all traffic demands. Figure~\ref{fig:drlEnvMidd} shows a schematic representation of the optimization scenario where the DRL agent interacts with the environment. This figure also illustrates how the DRL agent changes the routing policy of a traffic demand.

Traffic demands do not expire during the episode. This implies a challenging task for the DRL agent, since it has not only to identify critical resources on networks (e.g., potential bottlenecks), but also to plan ahead and optimize considering future demands. We summarize below the main features of the SDN-based routing optimization scenario considered in this paper:

\begin{itemize}
    \item The DRL agent makes sequential routing decisions for every traffic demand.
    \item It selects per-demand routing paths in order to minimize the most loaded link.
    \item Once a routing decision is made for a traffic demand, it cannot be re-routed, at the demand remains until the end of the episode.
\end{itemize}

\section{DRL+GNN routing optimization solution}
\label{sec:drl-agent}

We implement a DRL-based solution for routing optimization that integrates a GNN. We adapt the DRL+GNN solution from~\cite{almasan2019deep} for solving the routing problem addressed in the present paper (Sec.~\ref{sec:rout-optim-scenario}). \bbm{While the solution from~\cite{almasan2019deep} is tailored for routing optimization in optical networks, in this work we design a DRL+GNN agent that is tailored for Traffic Engineering optimization in IP networks.} 
Our agent implements the Proximal Policy Optimization (PPO) algorithm~\cite{schulman2017proximal}, which is a DRL on-policy algorithm based on actor-critic methods, in difference with the off-policy DQN algorithm used in the previous work. 
We also implement a simulation environment based on a fluid model, as the one used in~\cite{hartert2015declarative}. 

The DRL agent's training process is based on a trial-and-error process. At each time step, the agent receives a network state observation from the simulation environment, including the current link's utilization. \bbm{Then, the DRL agent uses a GNN to} construct an internal graph representation with the topology links. Each link has associated a state represented by a fixed-size vector with some real values (Sec.~\ref{subsec:environment}). With this representation, a message passing algorithm runs between the graph elements (i.e., the links of the network topology) according to the topology structure, and updates the link states \cite{gilmer2017neural}.

The outputs of this algorithm (i.e., the new links states) are aggregated into a global state, which encodes information about the whole network, and this global embedding is then processed by a fully-connected NN. Particularly, the GNN outputs a set of probabilities over the possible actions of the DRL agent. Finally, the output probability distribution is used by the DRL agent to sample the action to perform. 

\subsection{Network Simulation Environment}
\label{subsec:environment}

The network state is defined by features on the links, which includes the link capacity and the current utilization. \bbm{These features are stored in a fixed-size vector padded with zeros.} At the beginning of a DRL episode, all demands are allocated according to the OSPF routing (i.e., shortest path policy). The links states change as the DRL re-allocates traffic demands to specific \textit{src-dst} paths (i.e., sequences of links in the network). The final goal is to minimize the maximum link utilization at the end of the episode (i.e., when the DRL agent has iterated over all the traffic demands).

\subsection{Action space}
\label{subsec:action-set}

The number of possible routing combinations for each traffic demand (i.e., \textit{src-dst} node pairs) results in a high dimensional state/action space, even in small networks~\cite{suarez2019routingJ}. This makes the routing problem complex for the DRL agent, since it should estimate for each action which is the one that will lead to lower maximum utilization in the long-term. In other words, which is the optimal routing configuration per traffic demand that leads to the minimum maximum link utilization. Moreover, to exploit the generalization capabilities of GNNs over graphs, we need to define the action in a way that is invariant to edge and node permutation, using only link-level features rather than specific identifiers or labels. 

\begin{figure}[!t]
  \centering
  \includegraphics[width=0.7\linewidth,height=4.1cm]{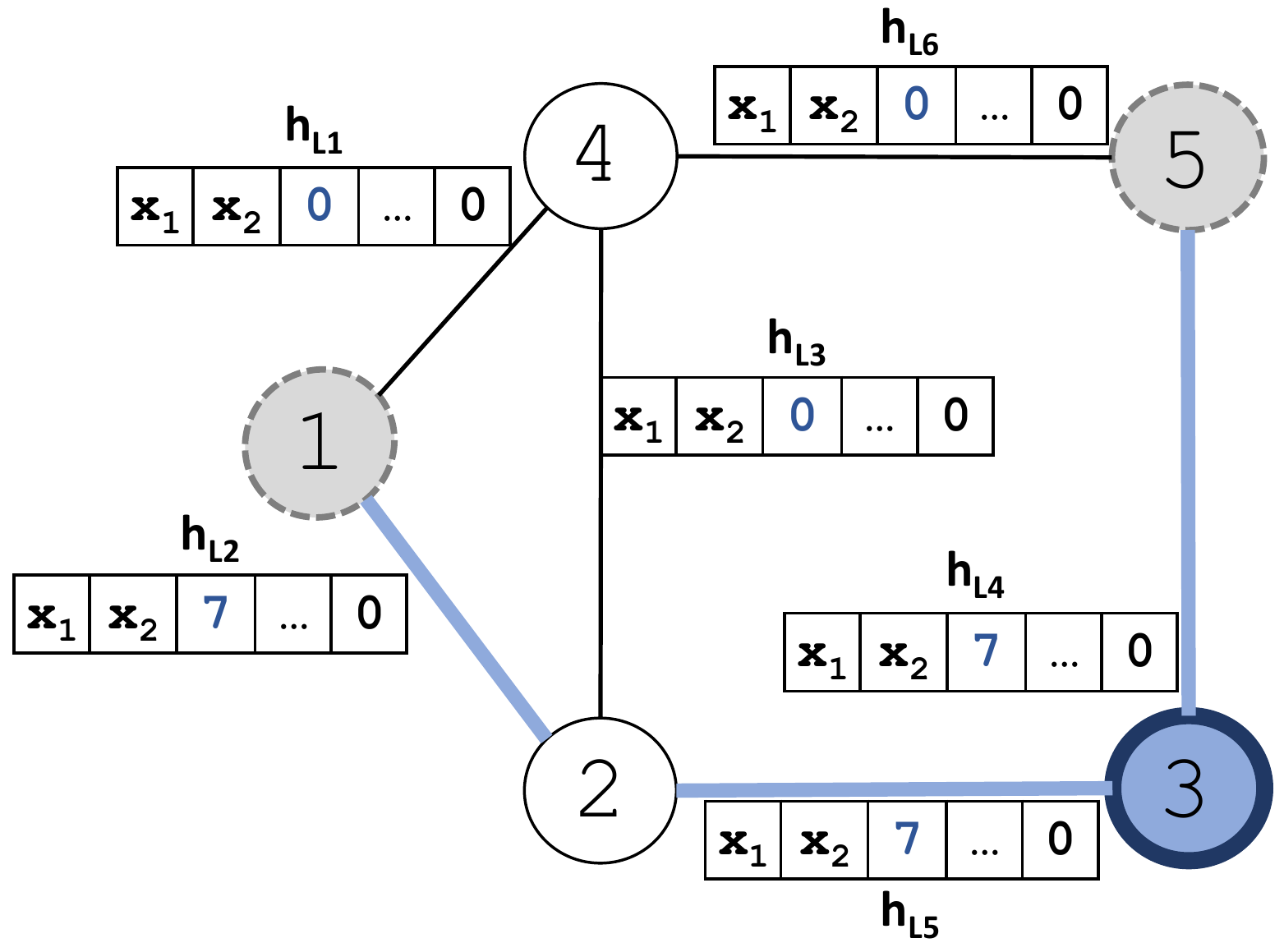}
  \caption{Action representation in the link states. Nodes 1 and 5 are the source and destination nodes respectively, and node~3 is the middlepoint selected.}
  \label{fig:actionspace}
  \vspace{-0.4cm}
\end{figure}

To overcome these problems, we use the middlepoint routing (MR) model \cite{hartert2015declarative} to define the action space. This model abstracts the routing paths between nodes as middlepoints, where the paths are pre-configured. In our paper, the pre-configured paths between nodes and middlepoints are computed using the OSPF routing policy. Given a traffic demand with an associated \textit{src-dst} node pair, we have as many different actions as number of middlepoints. Particularly, we limited the options to selecting only one middlepoint. This means that to reach a destination node \textit{dst}, the traffic demand can only cross at most one middlepoint before reaching the destination node. In our DRL agent, all the nodes except the source node can be middlepoint nodes, and \textit{src-dst} traffic demands can be directed through specific middlepoints using Segment Routing, as in~\cite{hartert2015declarative}.

To represent the action, we add an additional feature in the link's state (see Figure~\ref{fig:actionspace}). This feature includes the traffic volume of the current traffic demand $bw$ in case the link is within the selected path. Otherwise, the links that are not included in the path have this feature set to zero. Features \textit{x\textsubscript{1}} and \textit{x\textsubscript{2}} from Figure~\ref{fig:actionspace} are the link capacity and link utilization respectively. In this example, the traffic to allocate of the current demand is \textit{7}. 

\subsection{GNN architecture}
\label{subsec:gnnarchi}

\begin{figure}[!t]
  \centering
  \includegraphics[width=0.86\linewidth]{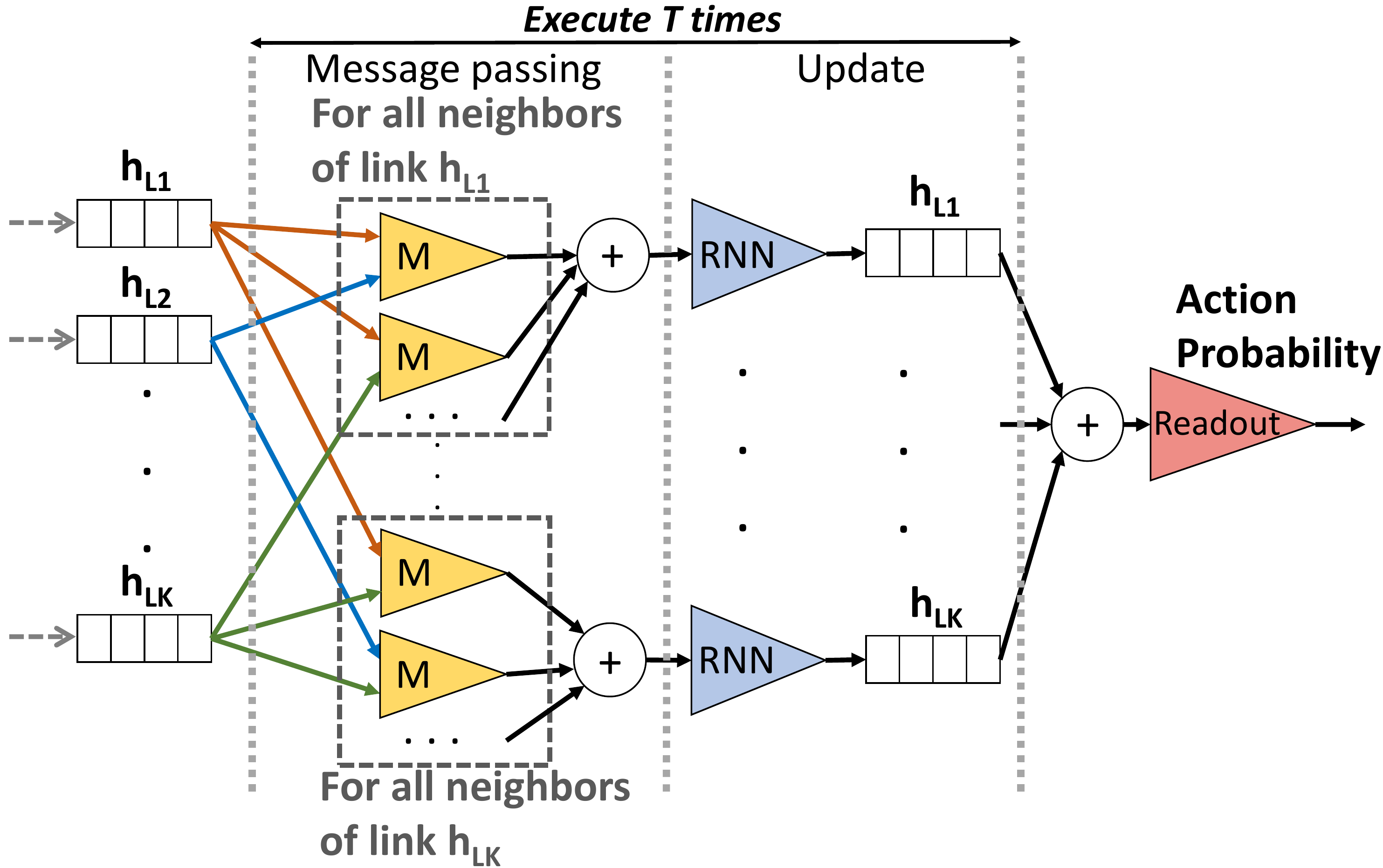}
  \caption{Message passing architecture.}
  \label{fig:mp_ppo}
  \vspace{-0.47cm}
\end{figure}

The GNN we implemented is based on the Message Passing Neural Network model \cite{gilmer2017neural}. In our case, we consider the network links as the graph entities and perform an iterative message passing process between adjacent links in the topology. \bbm{This process consists of combining the state of all links with those of their neighbours in the graph. Particularly, the states of connected links are processed by a fully-connected NN. Then, the results (called \textit{messages}) are aggregated for each link using an element-wise summation. Finally, the links' hidden states are individually updated using a Recurrent Neural Network (RNN), whose input is the previous link state and the new aggregated message.} This message passing process is performed \textit{T} times and, at the end of this phase, the resulting link states are aggregated using an element-wise summation. The result is then \bbm{a general graph embedding that is} passed through a fully-connected NN, which models the Readout function of the GNN. The output of this latter function is the action probability. Thus, given a traffic demand, this process is executed for each action (or possible middlepoint) and the final action of the DRL agent is sampled over the individual actions' probabilities. Figure~\ref{fig:mp_ppo} represents the internal architecture of the GNN, including the iterative message passing phase and the final readout.

\subsection{DRL Agent operation}

The DRL agent operates by interacting with the network simulation environment (Sec.~\ref{subsec:environment}), and the learning process is implemented using PPO \cite{schulman2017proximal}. 
First, we collect experiences from the DRL agent interaction with the environment. Then, using the collected experiences, we improve the current policy weights using mini-batch updates. Afterwards, the old experiences are removed and we collect new experiences using the new policy weights. This process is repeated until convergence or a pre-defined criteria is reached.

At the beginning of an episode, all traffic demands are allocated using OSPF routing. Then, the DRL agent iterates over all traffic demands and tries to change their current routing configuration. For each traffic demand, the DRL agent evaluates the action of re-allocating the demand on each possible middlepoint \bbm{using the GNN}. Notice that each middlepoint corresponds to an action and the traffic demand cannot be split between different middlepoints. Then, the GNN outputs a probability over each action, and the action performed by the agent is sampled from the resulting probability distribution of all actions. The chosen action is then applied to the network, leading to a new state (i.e., new links utilization), a reward, and a flag indicating if the agent finished iterating over all traffic demands. We set the reward to be the difference in the maximum link utilization between two steps. 

\section{Experimental Results}

\begin{figure}[!t]
    \begin{subfigure}[]{0.495\columnwidth}
	\includegraphics[width=1.0\linewidth,height=3.3cm]{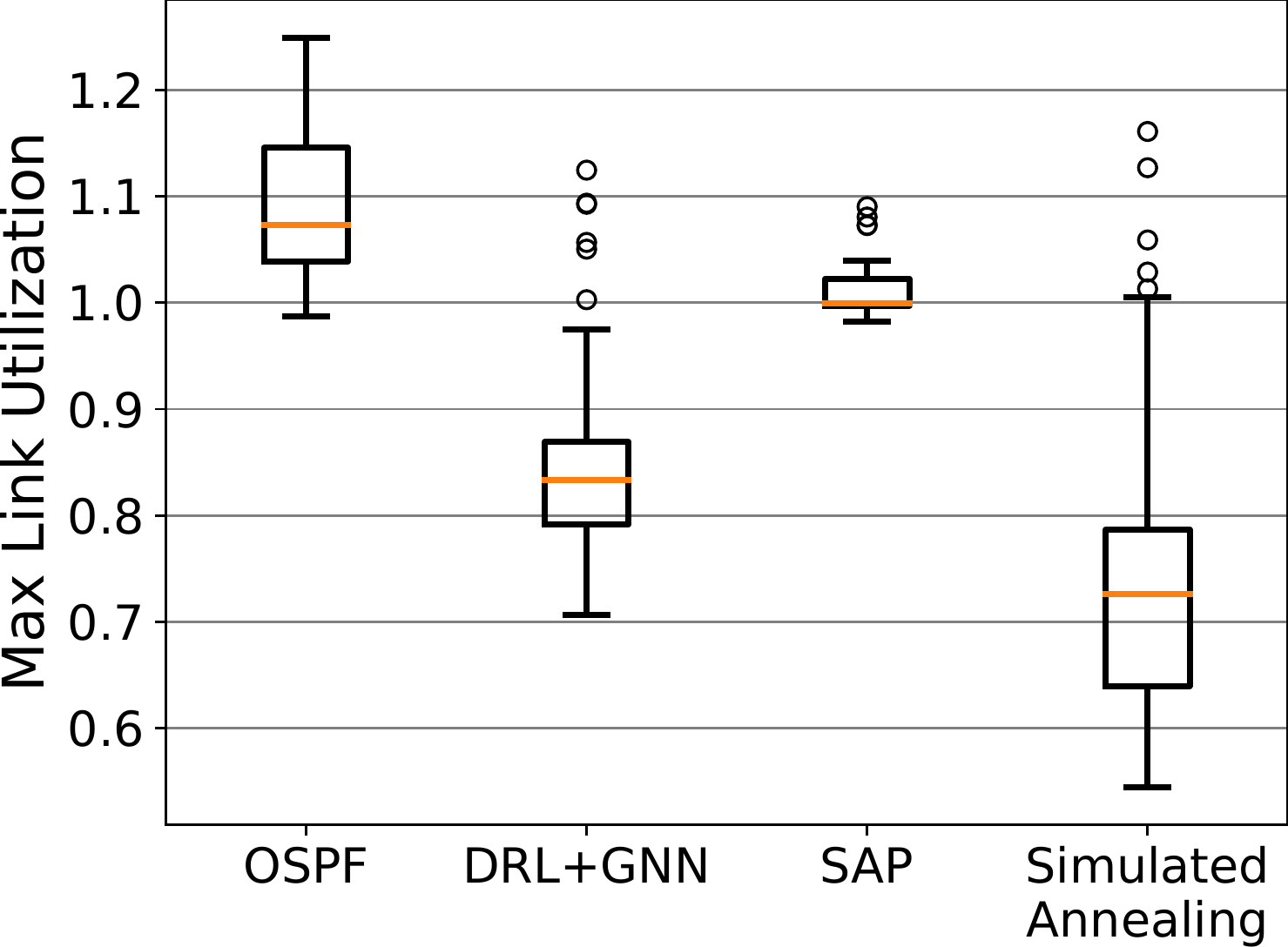}
        \caption{Evaluation on GBN}
	\label{subfig:perfevalGBN}
    \end{subfigure}
    \begin{subfigure}[]{0.495\columnwidth}
	\includegraphics[width=1.0\linewidth,height=3.3cm]{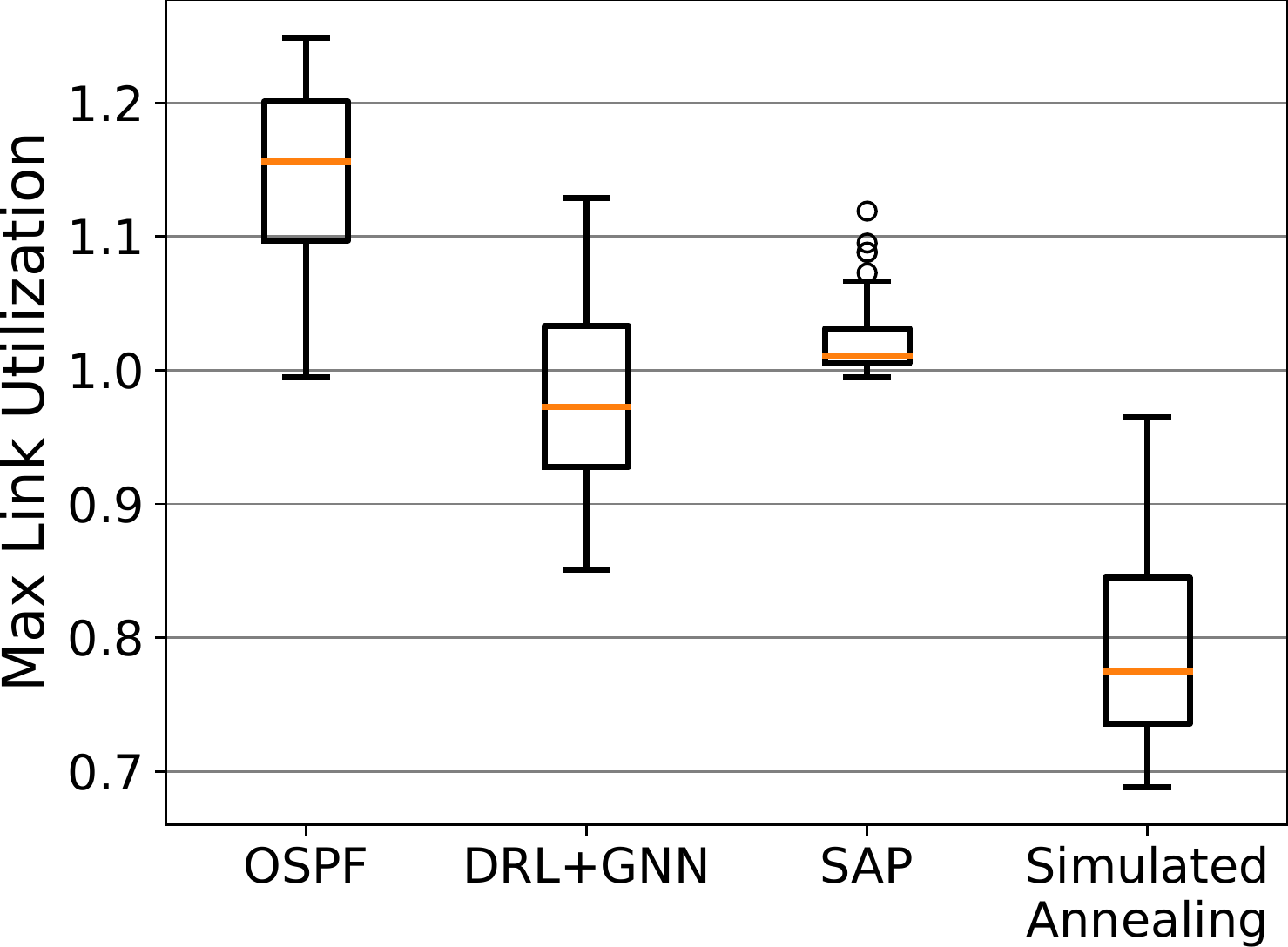}
        \caption{Evaluation on GEANT2}
	\label{subfig:perfevalGeant2}
    \end{subfigure}
    
     \medskip
    
    \begin{subfigure}[]{0.495\columnwidth}
	\includegraphics[width=1.0\linewidth,height=3.3cm]{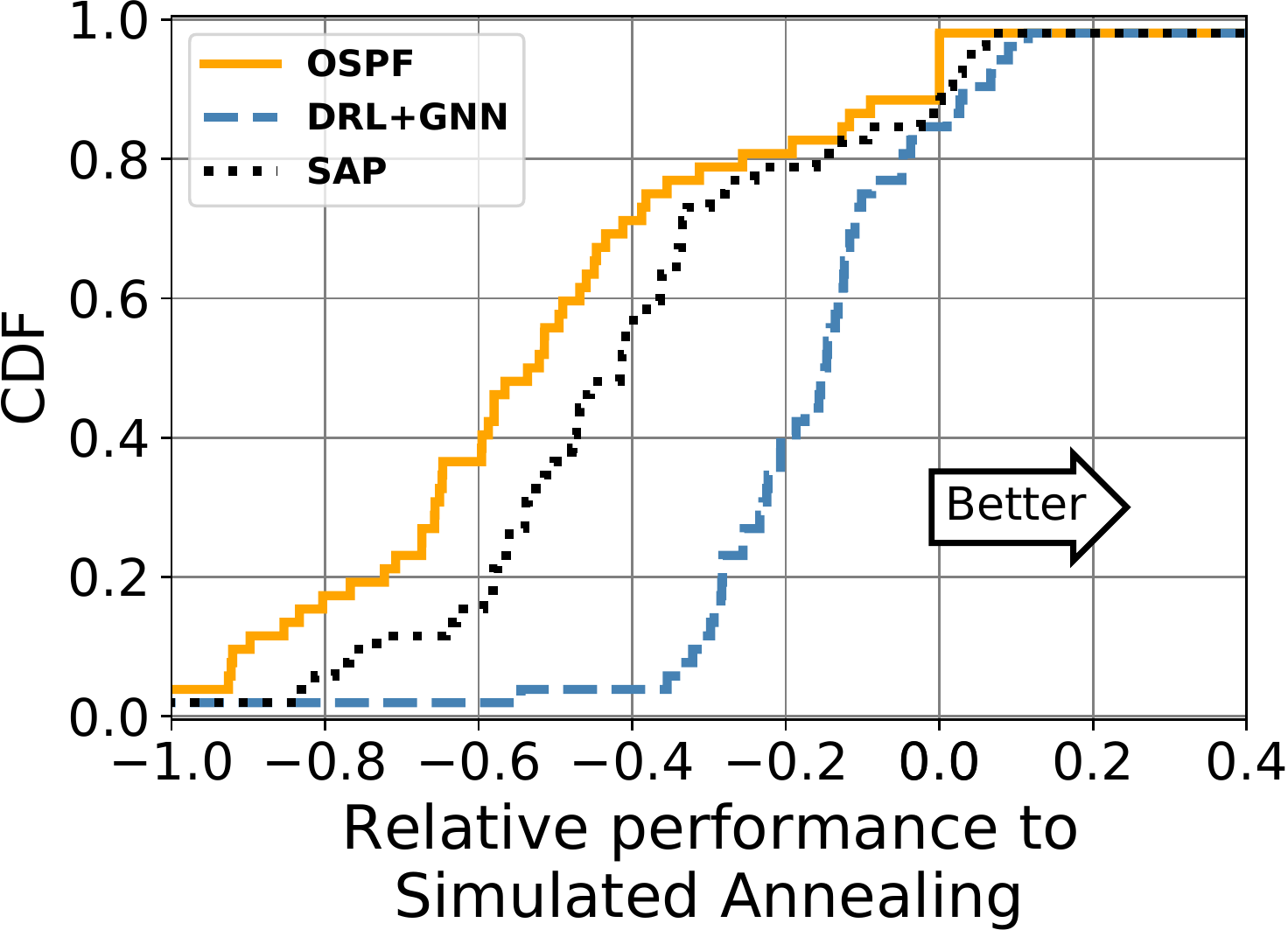}
        \caption{Evaluation on GBN}
	\label{subfig:perfcdfGBN}
    \end{subfigure}
    \begin{subfigure}[]{0.495\columnwidth}
	\includegraphics[width=1.0\linewidth,height=3.3cm]{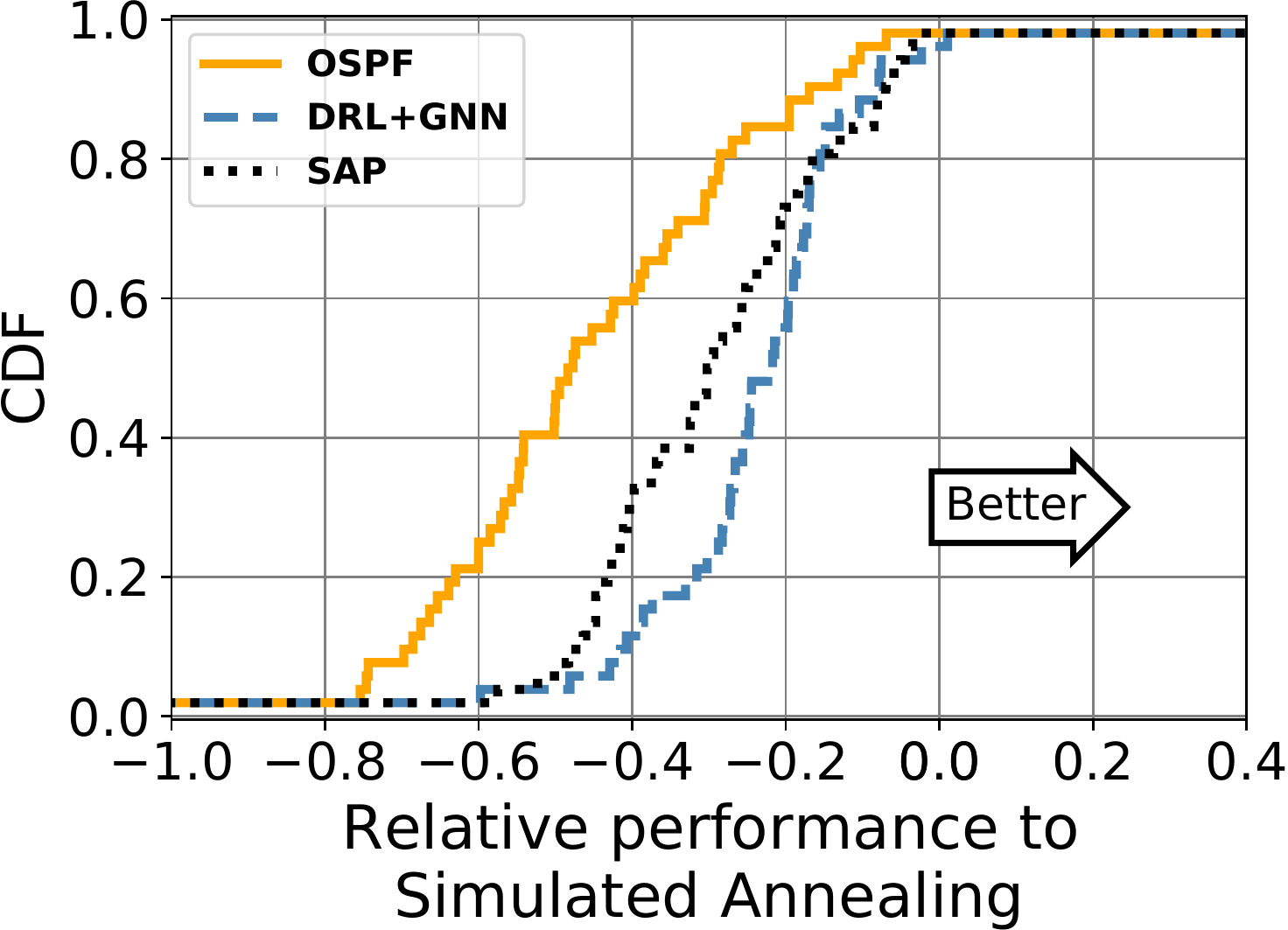}
        \caption{Evaluation on GEANT2}
	\label{subfig:perfcdfGeant2}
    \end{subfigure}
     \caption{DRL+GNN evaluation on GBN (a) and GEANT2 (b) using 50 TMs.}\label{fig:boxplots}
     \vspace{-0.5cm}
\end{figure}

\subsection{Evaluation setup}

We implemented the DRL+GNN solution of Section~\ref{sec:drl-agent} using Tensorflow~\cite{abadi2016tensorflow} and 
evaluated it on a network simulator implemented using the OpenAI Gym framework~\cite{1606.01540}. All the experiments were executed on \textit{off-the-shelf} hardware without any specific hardware accelerator (Ubuntu 20.04.1 LTS with processor AMD Ryzen 9 3950X 16-Core Processor). The simulation environment is configured by initializing the link features for each topology with the respective link capacities. All paths between nodes and middlepoints are pre-computed using OSPF, considering equal weights for all the links. The traffic demands are initially allocated following the pre-computed paths. Then, the environment sorts the demands in decreasing order of traffic volume, and the DRL agent sequentially re-routes them using MR.

Initial experiments were performed to choose an appropriate gradient-based optimization algorithm and hyperparameter values for the DRL+GNN architecture. We defined the links' hidden states $h_{l}$ as 20-element vectors (filled initially with the link features as shown in Figure~\ref{fig:actionspace}). In every execution of the GNN, we run \textit{T=5} message passing steps, and use mini-batches of 50 samples. The optimizer used is the Adam Optimizer~\cite{kingma2014adam}, with an initial learning rate of $2$·$10^{-4}$ and following an exponential learning rate decay during training. 

\subsection{Methodology}
\label{sec:method}

In our experiments, we use the NSFNet \cite{hei2004wavelength}, GBN \cite{pedro2011performance} and GEANT2 \cite{geant2} topologies. The traffic demands were generated using a realistic gravity model from \cite{roughan2005simplifying}. We consider four possible different values for the link capacities for the NSFNet topology and four different ones for the GBN and GEANT2 topologies. Specifically, the NSFNet has 5K, 10K, 15K and 20K as possible link capacities, and GBN and GEANT2 have 25K, 50K, 75K and 100K possible link capacities. We set the capacity individually on each link to ensure that the maximum link utilization is between 0.95 and 1.3 when applying the OSPF routing policy.

We compare the DRL's agent performance against two baselines. The first one is the Shortest Available Path (SAP). This consists on starting with an empty network, and iterate over the demands to allocate them on the shortest path that has more available capacity~\cite{suarez2019routingJ}. The second baseline we implemented is based on the Simulated Annealing  \cite{kirkpatrick1983optimization} algorithm. Starting from an initial routing configuration using OSPF, the algorithm performs an iterative process trying to decrease the initial state energy (i.e., minimize the maximum link utilization). In each step, the algorithm explores all possible next states and it decides whether to move to a new state or stay in the same using a temperature parameter. This parameter will guide the algorithm to move to states with lower maximum link utilization. As more iterations the algorithm performs, more chances it will have to approach to the global optimum. Therefore, we parameterized the algorithm to perform $4$·$10^{6}$ steps. This makes Simulated Annealing not suitable for real-time operation since it's cost (in time) is too high for finding good routing configurations. Specifically, Simulated Annealing spent around 30 minutes and 24 minutes for optimizing the GEANT2 and GBN topologies respectively.

We trained the DRL+GNN agent solely on the 14-node NSFNet topology \cite{hei2004wavelength}. During the training process, we used 100 different Traffic Matrices (TM) and we evaluated the agent performance on 25 TMs never seen during training. 

\subsection{Evaluation}

In the evaluation experiments, we compare our DRL+GNN agent against the baseline solutions. Particularly, we pick the model with more performance during the training phase, and evaluate it in two real-world network topologies: GBN (17 nodes) and GEANT2 (24 nodes). We used 50 different TMs to evaluate the performance for each topology.  

In Figures \ref{subfig:perfevalGBN} and \ref{subfig:perfevalGeant2} we can observe the evaluation results on both networks. Particularly, these figures show the maximum link utilization for the different optimization strategies. The OSPF label corresponds to the initial network state before optimization (i.e., shortest path routing). Likewise, in Figures \ref{subfig:perfcdfGBN} and \ref{subfig:perfcdfGeant2} we plot the CDFs of the previous experiments, but in this case in relative values w.r.t. the Simulated Annealing baseline. Note that Simulated Annealing represents a near-optimal solution that is not suitable for real-time operation as it has a very high computational cost. 

The experimental results indicate that our DRL+GNN agent outperforms the baseline heuristics (OSPF and SAP) on topologies not seen during the training phase (17 and 24 nodes), after being trained in a single 14-node topology. Specifically, the DRL+GNN agent reduces the maximum link utilization in average by $\approx$21.7~\% and $\approx$14~\% w.r.t. OSPF in the GBN and GEANT2 topologies respectively, while SAP only reduces it by $\approx$6.8~\% and $\approx$10~\%. This reveals the ability of our our DRL+GNN solution to adapt to different network scenarios -- even to new network topologies --, which is a essential property to operate in highly dynamic networks (e.g., traffic changes, link failures, new nodes), such as vehicular networks~\cite{zeadally2012vehicular}. In addition, the DRL+GNN agent was able to operate efficiently in scenarios with different link capacities than in the training scenario. This indicates that the DRL agent is able to adapt to different network topologies with different link features than those seen during training.

We also measured the average optimization time that each baseline spends on optimizing all the traffic demands. For the GBN topology we obtained an average time (in seconds) of $\approx$1476~s for the Simulated Annealing, $\approx$4~s for the DRL+GNN architecture and $\approx$0.05~s for the SAP heuristic. The optimization times for the GEANT2 topology are $\approx$1873~s for the Simulated Annealing, $\approx$12~s for the DRL+GNN and $\approx$0.15~s for the SAP. Notice that the larger is the topology (in number of nodes), the more traffic demands need to be optimized. This makes the problem complexity grow with the topology size.
However, DRL is inherently a technology based on NNs whose operation process can be easily parallelized using commodity hardware accelerators (e.g., GPUs). In consequence, DRL is key on enabling real-time network optimization.

\section{Discussion and 
Open Challenges}

In this section we discuss which are the main challenges that need to be addressed to enable the operation of DRL on dynamic networks. We consider that the most relevant aspects to address are generalization, action space design, and training cost.

\subsubsection{Generalization}
We argue that \textit{generalization} is an essential property for the successful adoption of DRL technologies in networks with dynamic topologies. In this context, generalization refers to the ability of the DRL agent to adapt to new network scenarios not seen during training (e.g., network topologies, traffic, configurations). With  generalization,  a  DRL  agent  can  be  trained  with a representative set of network topologies and configurations, and afterwards be applied to other scenarios never seen before.

In our work, we addressed generalization by designing a GNN-based DRL agent. In our experimental results, we have seen that our DRL+GNN agent is able to operate successfully in network topologies never seen before. From a commercialization standpoint, such ``universal'' DRL agent can be trained in a laboratory and later on be incorporated in a product or a network device (e.g., router, load balancer). The resulting solution would be ready to be deployed in a production network, without requiring any further training or instrumentation of the network.

\subsubsection{Action space} 
The definition of the DRL agent's \textit{action space} is fundamental to achieve a good optimization solution. An inadequate action space might explode as the topology size grows, having a direct impact in the complexity of the optimization problem to be addressed. Thus, it is important to find good network abstractions that can limit the dimension of the action space in order to facilitate the learning process to the DRL agent. In addition, the action space should offer enough flexibility and expressiveness to enable the finding of good routing configurations that lead to a high optimization performance.

In our work we leveraged MR \cite{hartert2015declarative} to limit the action space dimension (see Sec.~\ref{subsec:action-set}). We consider that MR offers a good level of abstraction to represent routing, but other methods can be used to define DRL's agent action space. For each traffic demand, we have as many actions as possible middlepoints. This means that the maximum number of actions are \textit{N-1}, where \textit{N} is the number of nodes (i.e., the action space grows linearly with the topology size). At the same time, this abstraction provides enough flexibility to achieve a near-optimal performance, as in~\cite{hartert2015declarative}.

\subsubsection{Training cost}
The \textit{training cost} of the DRL agent must be taken into account, especially when it is trained in large topologies. The DRL training process is, by definition, extremely sequential. This significantly hinders the parallelization of the learning process. New methods should be explored to scale DRL solutions to larger optimization scenarios. For example, these methods could be to accelerate the training process \cite{stooke2018accelerated}, transfer learning \cite{glatt2016towards}, or problem reduction \cite{sun2020scalable}.

Besides the challenges mentioned previously, there are some open challenges that require further attention by the research community and here we mention some of them. One of the open challenges is that DRL doesn't offer any kind of performance bounds. In other words, once a DRL agent is trained, there is no way to know exactly which are the upper or lower bounds in optimization performance. However, this problem is also present in many existing heuristic-based solutions.

\section{Conclusion}

In this paper, we explored a DRL agent based on GNNs for solving a routing optimization problem. To do this, we trained a DRL agent in a single topology and we evaluated it on two topologies never seen before. The experimental results show that the DRL agent is able to operate unseen network configurations while still obtaining a high performance. These results indicate that the DRL agent is able to adapt to different network topologies, which is an essential property to operate highly dynamic networks. In addition, the DRL agent has a low optimization cost, in the scale of seconds, and it's performance is comparable to near-optimal iterative solutions. This makes DRL a key technology to enable modern network applications that require real-time efficient operation in highly dynamic network topologies.

\section*{Acknowledgment}

This work has received funding from the European Union’s Horizon 2020 research and innovation programme within the framework of the NGI-POINTER Project funded under grant agreement No 871528. This paper reflects only the author's view; the European Commission is not responsible for any use that may be made of the information it contains. This work was also supported by the Spanish MINECO under contract TEC2017-90034-C2-1-R (ALLIANCE), the Catalan Institution for Research and Advanced Studies (ICREA) and the  Secretariat for Universities and  Research of the Ministry of Business  and  Knowledge of the Government of Catalonia and the European Social Fund.



%
\bibliographystyle{IEEEtran}
\bibliography{references}

\end{document}